# Analysis of Cell Packing Behavior to Enhance Wound Assessment

*Erin Kim*
*Phillips Academy, 180 Main St, Andover, MA 01810, USA*
*Guo Lab of Cell Mechanics and Department of Mechanical Engineering,*
*Massachusetts Institute of Technology, 77 Massachusetts Ave, Cambridge, MA 02139, USA*

**Abstract**

Wound assessment is a critical aspect of wound treatment, as the healing progress of a wound determines the optimal approach to care. However, the heterogeneity of burn wounds often complicates wound assessment, causing inaccurate wound evaluation and ineffective treatment. Traditional wound assessment methods such as Gross Area Reduction (GAR) and Percentage Area Reduction (PAR) are prone to misinterpretation, due to irregular results. Inaccurate wound assessment leads to higher rates of death and life-long physical and psychological morbidities in burn patients, especially in low-income communities that lack specialty care and medical resources. Therefore, I propose a novel approach to wound assessment: wound healing from the biophysical perspective of collective cell migration by analyzing cell packing behavior. This approach was modeled through Voronoi Tessellation simulations and applied to a wound healing system, where changes in the cell morphology parameters of aspect ratio and shape index were plotted over time to numerically evaluate the geometry of different cell migration packing patterns. Experimental results demonstrate the effectiveness of measuring aspect ratio, as a reduction in aspect ratio indicates that cell shapes become increasingly rounded throughout wound closure. This is further proven when considering physical principles in wound healing and changes in cell elongation. By placing a microscope objective on a phone camera, it is possible to directly examine any wound, with the calculations done on the phone as well. This efficient and accurate mechanism can be especially useful in low-resource communities, as it is accessible regardless of technical or medical background.

# 1 INTRODUCTION

## 1.1 Burns

Burns are identified as a global public health problem by the World Health Organization, accounting for approximately 180,000 deaths annually, with increased frequency and severity of burns associated with lower educational status, lower income, and substandard living conditions [World Health Organization, 2018]. Almost two-thirds of total deaths occur in African and Southeast Asian countries due to insufficient awareness of preventative measures, increased exposure to environmental hazards, and lack of accurate wound assessment leading to ineffective treatment. In particular, the current system of burn care delivery is hampered primarily by the growing shortage of burn-specializing physicians and disparities in access to medical information [Delaplain and Joe, 2018]. Despite burn wound outcomes defined by countless racial, ethnic, socioeconomic, and educational disparities, burns are widely under-appreciated and burns' immense heterogeneity remains relatively unknown by the general public.

Burn injuries can be caused by a number of reasons, including friction, cold, heat, radiation, chemical, or electric sources, but the majority of burns are caused by heat from hot liquids, solids, or fire [Laurent et al., 2020]. There are several classifications for burn wounds according to severity, defined by the wound's depth and size [Victorian Adult Burns Service, 2022]. First-degree burns affect the epidermis, the uppermost layer of the skin; the skin reddens and pain is limited in duration. Second-degree burns of superficial partial-thickness—although there is the possibility of scarring—do not require surgery but need wound care and dressing. Second-degree burns that have deep partial-thickness are drier and require surgery, but are relatively less painful, due to the partial destruction of the pain receptors. Third-degree burns are full-thickness and extend through the entire dermis, the underlying skin layer. While not typically painful as the depth of the burn damages nerve endings, third-degree burns require intensive care and protection from becoming infected. Finally, fourth-degree burns involve injury to deeper tissues, such as muscle or bone, which often become blackened and frequently lead to loss of the burned portion. Burns are also categorized by size into minor or major burns. A minor burn encompasses <10% of the total body surface area (TBSA). >10% TBSA in elderly patients,

>20% TBSA in adults, and >30% TBSA in children are widely considered major burn injuries [Jeschke et al., 2020].

## 1.2 Wound Healing

The wound healing process, which takes months to years as the scar undergoes remodeling, consists of four main phases: hemostasis, inflammation, proliferation, and remodeling [Falanga, 2005]. In order to provide ongoing individualized and efficient treatment, wound assessment by tracking a patient's status throughout these healing phases is essential. While failure to progress in the stages of wound healing can result in chronic wounds, careful wound care based on a thorough understanding of the wound can streamline the stages of healing by keeping wounds moist, clean, and protected from reinjury and infection. However, the wound healing phases, despite being linear, are continuous and oftentimes overlap. Furthermore, the phases can progress backward or forward depending on internal and external patient conditions. These commonly occurring deviations obscure physicians' assessment of a patient's precise wound healing progress.

The heterogeneity of severe burn wounds, in particular, can present unique challenges. As burn wounds generally take skin 12–18 months—a relatively long period for wound treatment—to finish healing and fade to a near-normal color [Regions Hospital, n.d.], wound assessment of evolving extent and severity is often challenging. Especially, burn wound assessment with the naked eye presents hardships, as it is difficult to differentiate healing from suffering an early-stage infection. In fact, burn wound infections are the most common complication and the leading cause of death in burn patients [Jeschke et al., 2020]. Untimely treatment can lead to worsened infection outcomes, degeneration to chronic wounds, life-long physical and psychological morbidities, or even death. Disability and death resulting from burn wounds are particularly prevalent in communities of low socioeconomic status due to the lack of accurate wound assessment leading to ineffective treatment.

To quicken the readout of the wound healing process and treat early signs of wound infections, I propose an approach to wound healing from the biophysical perspective of collective cell migration: using the cell morphology parameter, aspect ratio, to detect wound healing progression. This would allow early recognition of signs of wound infection and can help wound care experts to quickly intervene with treatment, improving burn patient outcomes.

## 2 BACKGROUND THEORY

### 2.1 Current Metrics for Wound Healing Measurement

Conventional approaches for wound assessment include Gross Area Reduction (GAR), which measures changes in raw area, and Percentage Area Reduction (PAR), which measures changes in area as a percentage of the initial area. In both methods, the wound is measured at several time intervals to establish its healing rate, which is quantified by the size of the wound measured through tracing or using photographic methods at a particular time point. Following the introduction of the intervention, the healing rate is reevaluated to detect any effects.

However, there are discrepancies in measurement since different wounds are considered as healing faster depending on the metric that is used. Clinical data indicates that GAR exaggerates large wounds, while PAR exaggerates small wounds. [Gorin et al., 1996] and [Bull et al., 2022] represent this point through two hypothetical wounds, A and B, simplified as circles for descriptive purposes. Wound A originally has a radius of 5 cm, and over the course of 4 weeks reduces to 4 cm. Throughout the same time period, Wound B reduces from a radius of 3 cm to 2 cm.

When adopting GAR as the metric, the initial area of Wound A is $\pi r^2 = 79\ cm^2$ and the final area of Wound A is $\pi r^2 = 50\ cm^2$, resulting in a GAR of $79\ cm^2 - 50\ cm^2 = 29\ cm^2$; the initial area of Wound B is $\pi r^2 = 28\ cm^2$ and the final area of Wound B is $\pi r^2 = 13\ cm^2$, resulting in a GAR of $28\ cm^2 - 13\ cm^2 = 15\ cm^2$. The use of GAR therefore leads to a conclusion that Wound A is healing faster than Wound B. On the other hand, when using PAR as the metric, the PAR of Wound A is defined by $\frac{100 \times (79\ cm^2 - 50 cm^2)}{79\ cm^2} = 37\%$. The PAR of Wound B is $\frac{100 \times (28\ cm^2 - 13 cm^2)}{28\ cm^2} = 53\%$, which leads to the conclusion that Wound B is healing faster than Wound A.

This abnormality is clarified by the quadratic relationship between radius and area (area is proportional to the radius squared), causing changes in area to be proportionally exaggerated in larger wounds for a given change in radius. On the contrary, PAR (in which change in area is divided by the original area) is proportionally minimized in larger wounds. Such inconsistencies give rise to major issues: the first being the domination of the healing of large wounds relative to the healing of small wounds. The second is that, over the course of wound size reduction throughout the study, early healing rate will be magnified relative to later healing rate, complicating longitudinal comparisons. Furthermore, both methods require having a whole image of the wound and at least two time points to measure the healing speed, which cannot always be met in application. Therefore, I propose utilizing the role of collective cell migration in wound healing by measuring cell shape, which changes as movement occurs. This new method would determine the wound healing stage by looking primarily at a local image of cells at one time point, which is easier and faster.

## 2.2     Correlation Between Collective Cell Migration and Cell Morphology

Collective cell migration is the coordinated movement of a cohesive group of cells that maintain their intercellular connections and collective polarity. This behavior is essential for physiological and pathological processes pertaining to the skin such as wound repair. During wound healing, epithelial cells proliferate and migrate collectively as a coherent sheet into the center of the wound. Cells move en masse because collective cell migration, being a "multicellular structure," enables cells to better respond to chemical and physical cues compared to isolated cells. As the primary goal of the epithelial cells is to restore the epithelial barrier, it is important for the cells to maintain proper cell-to-cell adhesion when migrating over the wound bed so that the epithelial barrier is not further compromised. Directional migration of the cell sheet, induced by the wounding of an epithelial monolayer, maintains this tight intercellular adhesion [Li et al., 2013]. Thus, studying the collective migration of cells in the context of wound healing allows the simulation and exploration of critical mechanisms of action involved in the process.

In collective cell migration, cell shape plays an essential role in the regulation of various migratory behaviors. Cells are thoroughly influenced by interactions with their neighbors while migrating, thereby inducing changes in cells' shapes and packing patterns. Cell shapes—given that they are products of the cell's material and active properties balanced by external forces—rely on the tight regulation of intracellular mechanics and the cell's physical interaction with its environment [Paluch and Heisenberg, 2009]. Cell morphology can be defined by the number of neighbors that each cell has. Using a simplified simulation model, the number of neighbors of cells can be counted, and packing patterns can be observed.

# 3     EXPERIMENTAL METHODS

## 3.1    Simulation of Particle Packing on Spheres

To further my understanding of cell shape and packing patterns, a computational analysis was performed on MATLAB_R2020a as a simplified model of the biological system. Particles, representing individual cells, were generated on curved surfaces in numbers ranging from 0 to 1,000, for the observation of particle packing patterns in relation to sphere curvature. Through this approach, many data points that collectively advanced my understanding of particle shape were produced—either hexagons or polygons with more or fewer borders. Statistical plots were then created using Voronoi tessellation, which allowed the analysis of cell shapes and cell-to-cell networks on the systemic level. Voronoi tessellation was done by drawing perpendicular bisectors from center points to the line joining two stations. Types of polygons and their frequencies based on cell count were plotted.

## 3.2    Study of Wound Healing Models

This theory can be applied to real-life cells—but cells in real life are not as defined as the particles simulated in Voronoi tessellation, and as a result, may be more difficult to analyze and track. Therefore, a more physiological system—a wound healing system—was utilized to further understand the changes in cell shape and packing patterns. The Java-based image processing program, ImageJ, was used to edit, process, and analyze images. To quantitatively define cell shape, this analysis employed methods of counting the number of neighboring cells, as well as using shape index—a cell morphology parameter that suggests unjamming or fluidization of cell packs—and aspect ratio—another cell morphology parameter that indicates how elongated and nonspherical the shape of a cell is—to numerically evaluate the variety of packing patterns that are formed as a result of cell migration.

A video of wound healing in live *Drosophila melanogaster* wing imaginal discs obtained from a confocal fluorescent microscope [Tetley et al., 2019] was modified and analyzed using ImageJ. After converting images to 8-bit, morphological segmentation was performed. Threshold and tolerance values were adjusted and watershed lines were used to segment grayscale images. Particle analysis was then conducted for each frame of the video, separated by increments of 10.

Shape changes, including aspect ratio and shape index during wound closure, were then analyzed, using cell characteristics such as area, perimeter, and solidity. Shape index and aspect ratio were each calculated using the following equations:

$$Shape\ Index\ p\ =\ \frac{perimeter}{\sqrt{area}}$$

$$Aspect\ Ratio\ AR\ =\ \frac{L_{max}}{L_{min}}$$

## 4 RESULTS

### 4.1 Polygon Fraction as a Function of Particle Number

Random particles were generated on a sphere where the particle number N = 11 (Figure 1a) and where N = 900 (Figure 1b). Polygons, Z, were divided into three categories: Z > 6, Z = 6, and Z < 6, indicated through the colors, blue, white, and red on the sidebar, respectively. As shown in Figure 1a, when N was a small number, red polygons with a number of sides less than 6 were the majority. On the other hand, when N was a large number, blue particles with a number of sides more than 6 made up the majority. After this quantitative analysis, a qualitative analysis of the average and standard deviation of polygon fraction and particle numbers was performed (Figure 2).

Polygon shape was dependent on particle number; polygon fractions varied in accordance with differing particle numbers. When the particle number was large, polygon shapes were independent of the particle number, with the majority of the shapes being hexagons. Therefore, the number of neighbors a cell has when packed together will apply to the cellular function in the curved surfaces.

Below is the visualization and analysis from running my code on MATLAB. Random points on a sphere were generated to be used as centers of Voronoi tessellation (Figure 1). The average and standard deviation of each particle number, N, were examined and the polygon fraction was plotted as a function of the particle number (Figure 2).

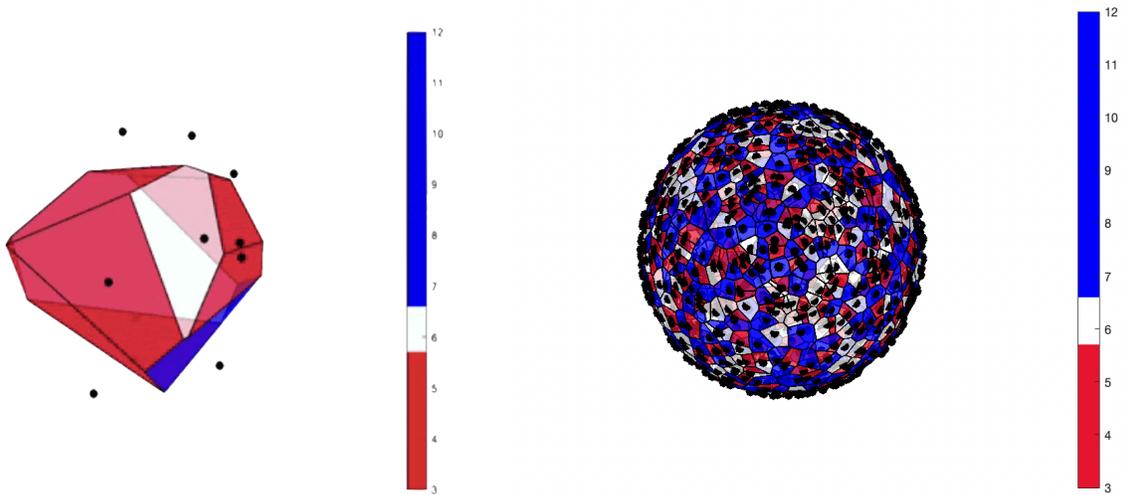

Figure 1 | Simulations generated from the number of particles.
a, The number of particles on the sphere is set to 11, with 10 frames (left).
b, The number of particles on the sphere is set to 900, with 10 frames (right).

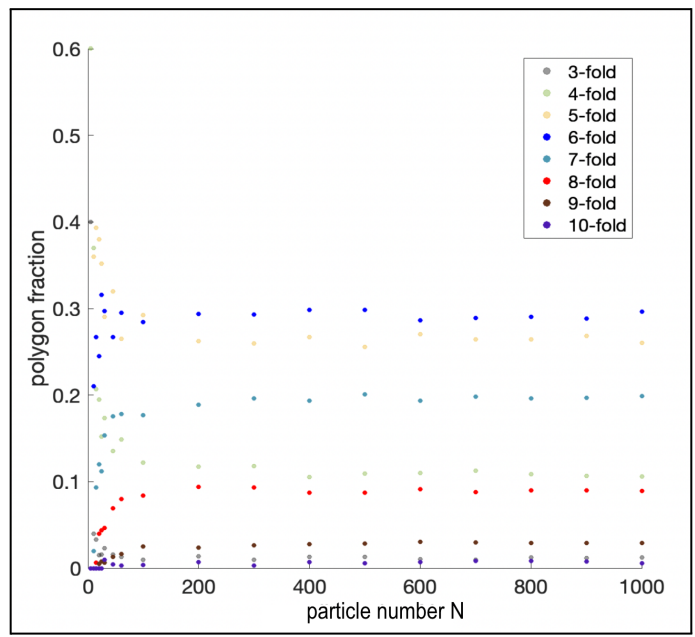

Figure 2 | The evolution of polygon fraction relative to the variation of particle number.

There is a clear distinction between polygon fractions when N is different. Figure 2 is an analysis of variations in polygon fraction trends throughout the particle number range, 0–1000. It is clearly shown that hexagons, indicated as "6-fold" in the legend, are the largest polygon fraction throughout the majority of particle numbers. Yet, as the particle number increases, there are some polygons whose fractions increase, such as octagons ("8-fold") and heptagons ("7-fold"), while others, like pentagons ("5-fold") and rectangles ("4-fold"), show a downward trend. These increasing and decreasing trends occur as the average vertex number outside of this model is determined by topological defects.

### 4.2 Aspect Ratio as a Metric of Wound Assessment

Given the applications of the study of cell migration in practical, medical conditions such as wound repair, ImageJ was used for a clearer analysis of real-life wound repair videos. Pictured below are specific frames in intervals of 10 in a wound repair video [Tetley et al., 2019] that were used (Figure 3) and processed through ImageJ by adjusting the threshold and tolerance values. Image segmentation was performed by dividing each figure into many pixels, with each pixel being a different color intensity. As the program was able to detect significant jumps in intensity, it distinguished the boundaries by differentiating the varying values of color intensities. The characteristics of the region of cells analyzed were then collected, with a dataset returned by ImageJ.

First, the average cell shape as a function of frame number (Figure 4) was measured. Assessing the cell shape of the first layer of cells around the wound, average cell shape showed a decreasing trend, along with a decrease in wound area (Figure 5). Upon checking the video, it was confirmed that there was no correlation between the wound area and the turning point of average cell shape—30 and 40 frames. Standard deviation increased with time, which suggested that cell shape distribution was more heterogeneous with time. Plots for the aspect ratio of the cells over time aligned with these results, revealing a decrease in aspect ratio over time as wound healing progressed (Figure 6). On the other hand, no significant change was observed for the shape index values (Figure 7).

| Frame Number: | 10 | 20 | 30 |
|---|---|---|---|
| Segmentation Figure: | 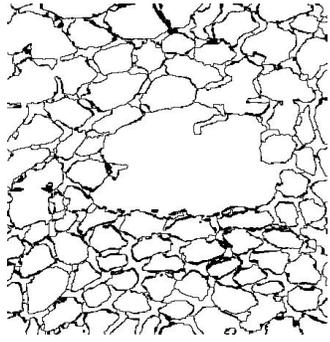 | 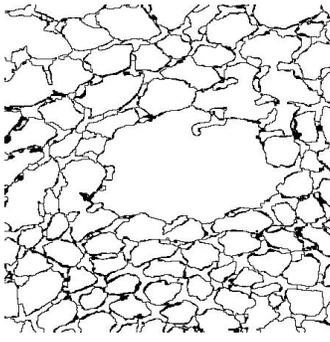 | 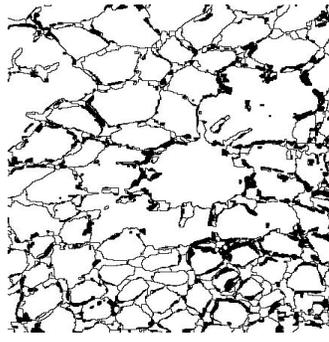 |
| Frame Number: | 40 | 50 | 60 |
| Segmentation Figure: | 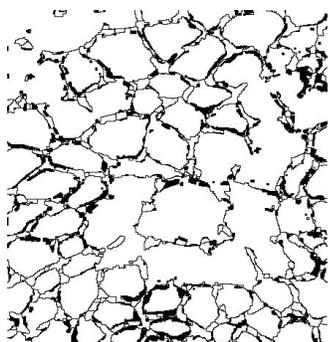 | 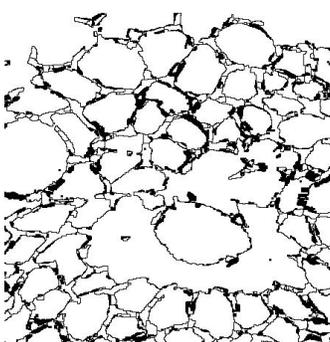 | 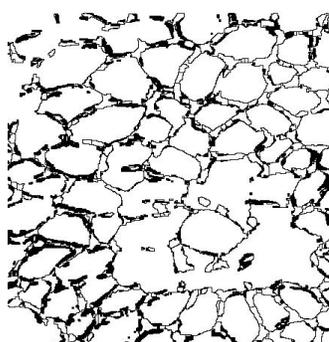 |

Figure 3 | Wound repair segmentation figures for each frame number.

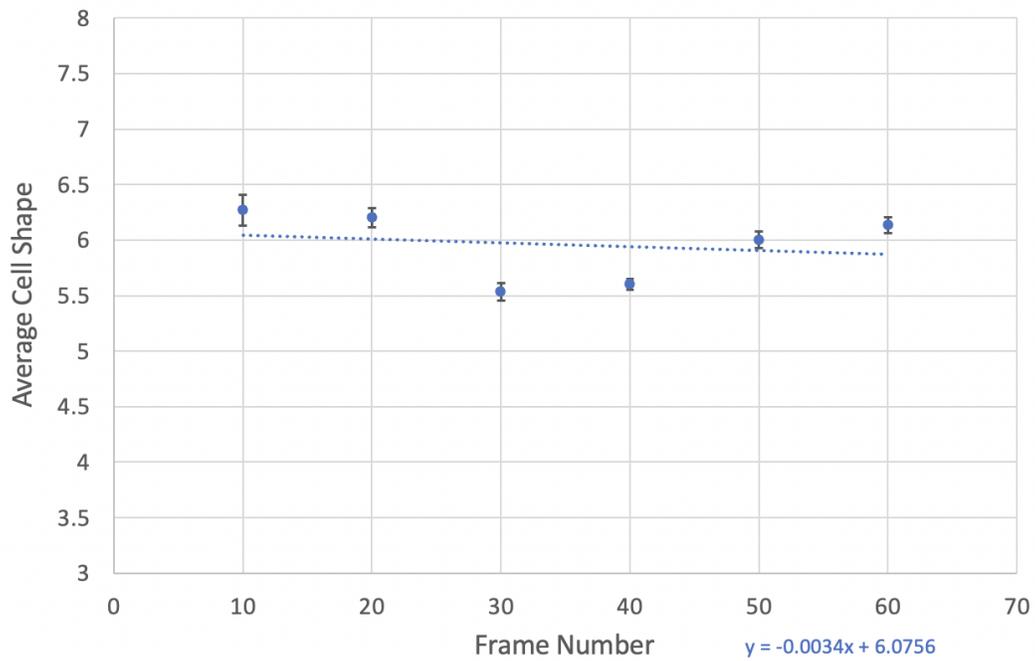

Figure 4 | Changes in Average Cell Shape for each cell over time.

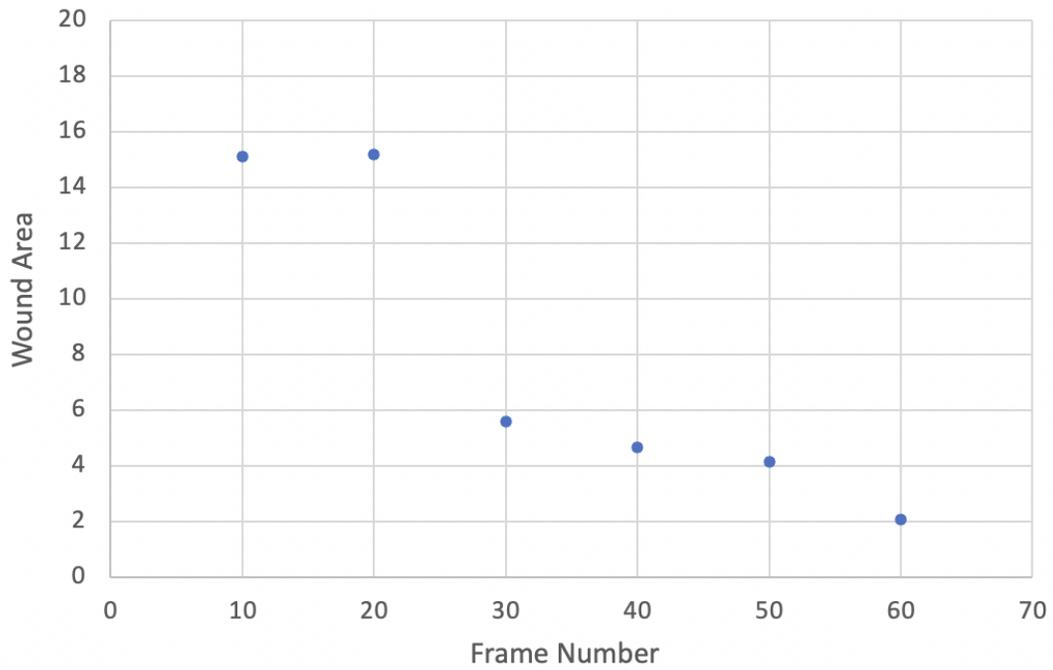

Figure 5 | Measurement of Wound Area for each Frame Number throughout the wound healing process.

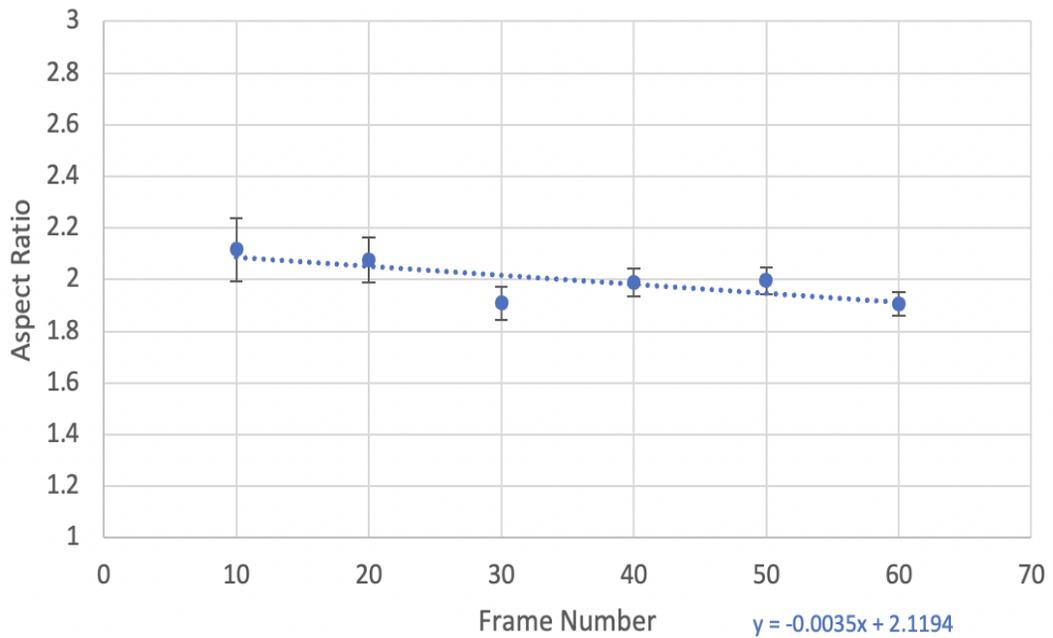

Figure 6 | Plot of local numbers: Changes in Aspect Ratio for each cell over time.

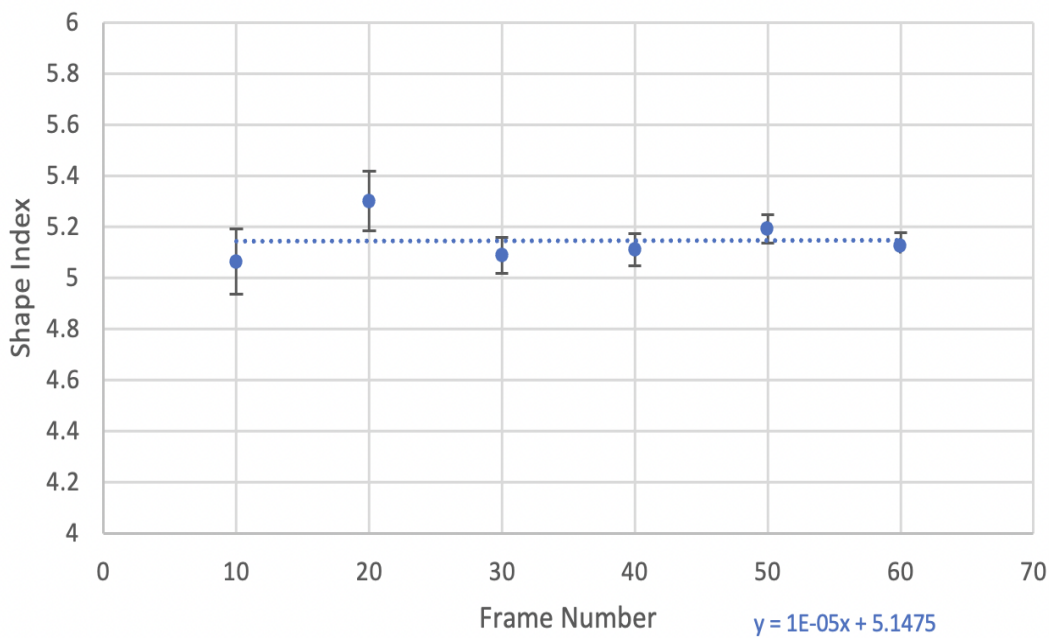

Figure 7 | Plot of local numbers: Changes in Shape Index for each cell over time.

# 5 CONCLUSION

Collective cell migration is highly dependent on cell morphology, making it useful to gain insight into physiological processes involving migratory behaviors, such as wound healing, by observing how cell shapes and packing patterns are altered at various time points. On the analysis of cell morphology, the prevalence of hexagons in the Voronoi tessellation plot (Figure 2) supports the hypothesis that out of the various polygonal shapes cells can assume, cells that have six neighbors indicate the most consistent structure. Thus, hexagons—which have six borders and six directly neighboring cells—are the most common cell shapes in an intact monolayer. The hexagons of honeycomb are a prominent example of such cell packing patterns that can be spotted in nature.

When epithelia or cell monolayer is wounded, cells need to rearrange themselves to close the wound. During this process of wound closure, cells elongate themselves to quicken their migration to close the wound. As the wound closes, the cell aspect ratio gradually decreases, as cells become more hexagonal to optimize arrangement over time. This is the fluid-solid transition—cell fluidity aids rearrangement while being solid means that the epithelia are back to their regular intact structure. All of this can easily be seen through simple microscopic imaging of cells through a single ImageJ and MATLAB analysis. These data would allow us to understand the extent to which a given wound has recovered and likely how much longer it would take for the full closure. With a standard curve characterizing aspect ratio change over the wound closure process, we would be able to have knowledge of where we are now and approximately how much longer healing will take just by examining the aspect ratio at any time point. This data would significantly deepen our understanding of wound healing on the cellular level, and it would allow physicians to track the status of the wound and project the likely outcome. Furthermore, by placing a microscope objective on one's phone camera, one can directly examine any wound and complete calculations using a single smartphone. This device could be conveniently used by anyone without a technical background, meaning it could drastically improve burn wound management and treatment, particularly in low-resource communities.

      The application to a wound healing system on ImageJ and the decrease in aspect ratio (Figure 6) would likely result from the elongation of cells as wound healing evolves over time. The cells' primary and initial goal during wound healing is to reach the center of the wound and fill the wound as soon as possible; therefore, cells are elongated to migrate faster. Cells with a higher aspect ratio are likely to migrate faster as they need the polarity to move directionally. However, as time passes and the wound closes, the cells slow down and become more rounded, resulting in a lower aspect ratio. The shape index, a relatively new concept that claims to suggest unjamming or fluidization of cell packs, did not display a significant change in the resulting plots (Figure 7). This indicates that perhaps the fluidity of cells does not change much during the wound healing process.

      Using the cell morphology parameter of aspect ratio to accurately evaluate the progress of wound healing, it is possible to provide optimized, individualized, and efficient wound treatment that will improve patient outcomes across the board.

# 6     ACKNOWLEDGEMENTS

I thank Prof. Ming Guo, Dr. Yulong Han, Wenhui Tang, and the other members of the MIT (Massachusetts Institute of Technology) Guo Lab of Cell Mechanics for their helpful advice on the manuscript. I also wish to acknowledge NSF EBICS (National Science Foundation Center for Emergent Behaviors of Integrated Cellular Systems) for generously providing all resources and funds necessary for experimentation in this study.